\documentclass[10pt,leqno]{amsart}
\usepackage{graphicx}
\usepackage{indentfirst,csquotes}

\usepackage{amssymb,amsthm,amsmath}
\usepackage{xcolor,paralist,hyperref,titlesec,fancyhdr,etoolbox}

\usepackage{hyphenat}

\titleformat{\section}[display]{\normalfont\huge\bfseries\centering}{\centering\chaptertitlename\thechapter}{10pt}{\Large}
\titlespacing*{\section}{0pt}{0ex}{0ex}

\hypersetup{ colorlinks=true, linkcolor=black, filecolor=black, urlcolor=black }

\usepackage{lipsum}

\begin{document}
\title[Constructing Fundamentals for the TPASAI]{Constructing Fundamentals for the Theory of Proportions and Symbolic Allusions Applied Interdisciplinarily} 
\author[D. Babuc]{Diogen Babuc}
\date{\today}
\address{Blvd. Vasile Pârvan nr. 4, Timişoara 300223, Timiş Romania}
\email{diogen.babuc@e-uvt.ro}
\maketitle

\let\thefootnote\relax

\begin{abstract}
The Theory of Proportions and Symbolic Allusions Applied Interdisciplinary (TPASAI) is a framework that integrates mathematics, linguistics, psychology, and game theory to uncover hidden patterns and proportions in reality. Its central idea is that numerical encoding of symbols, dates, and language can reveal recurring structures and connections that reflect universal principles. By applying fractal analysis, the theory identifies patterns across different scales, offering a unifying perspective on the structure of the world.
One key aspect of TPASAI is symbolic analysis, which allows for the reinterpretation of traumatic experiences in psychotherapy. For example, assigning numerical values to elements like fingers, dates, or words can help individuals uncover meaningful associations between personal experiences and collective symbols. This approach encourages cognitive flexibility and provides a therapeutic avenue for recontextualizing emotions and memories.
The theory also incorporates principles of game theory, which frame reality as a system of symbolic ``codes'' governed by rules that can be understood and strategically used. This perspective is especially useful for psychological conditions like obsessive-compulsive disorder (OCD), enabling patients to approach their obsessions as decipherable patterns rather than rigid constraints.
TPASAI has practical applications in psychology, education, and technology. In education, it aids in teaching mathematical and linguistic concepts by exploring connections between symbolic representations and real-world events. In technology, the methodology can be employed in ciphering and natural language processing through symbolic and proportional encoding.
The innovation of TPASAI lies in its ability to merge the structured rigor of mathematics with the interpretative flexibility of symbolic analysis, offering a deeper understanding of events and relationships. Its interdisciplinary nature opens new avenues for research and practical applications across various fields.
 
\end{abstract} 

\bigskip
\bigskip
\bigskip

\noindent \LARGE \textbf{I Introduction}
\\
\\
\noindent \large In exploring the complexity of interactions between symbols, numeration, and human perception, a symbolic theory emerges that combines elements from mathematics, psychology, and philosophy. This theory suggests that attributing numerical meanings to everyday entities, such as fingers or components of language, can facilitate a profound understanding of reality and provide therapeutic tools for individuals dealing with psychological disorders.

For example, the numbering of fingers is not merely an arithmetic exercise but also a symbolic process where each finger, numbered from 1 to 10, can represent calendar dates or significant moments. The third finger, for instance, can correspond to the third day of the month, and the tenth finger to October, together forming the date 3 October. Such associations are not arbitrary; in many cultures, numbers and dates carry symbolic meanings and can be correlated with personal or collective experiences, including traumas.

Additionally, analyzing words by breaking them into syllables and assigning numerical values to each part can reveal unexpected connections between language and human experience. For instance, the word ``suspect'' can be divided into the syllables ``sus--pect,'' each with specific lengths, whose concatenation (3, 4) could lead to the number 34. There can be multiple interpretations of this number. This approach highlights how numerological coding of language can reflect deep religious, psychological, or philosophical themes \cite{singer}.

\begin{figure}
    \centering
    \includegraphics[width=1\linewidth]{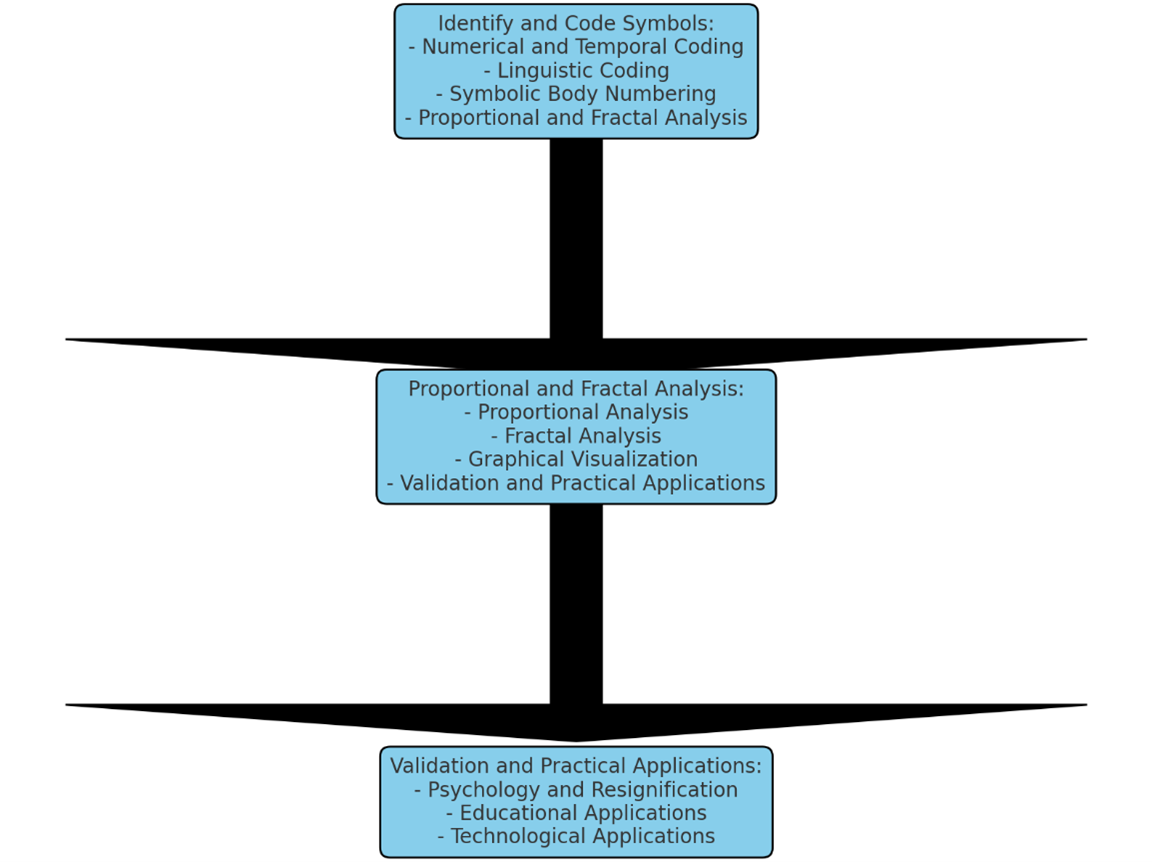}
    \caption{TPASAI methodology with the adjusted hierarchical layout.}
    \label{fig:enter}
\end{figure}

Applying such codifications and symbols to psychological disorders, such as OCD, paranoia, or schizophrenia, can offer patients a new perspective on reality. Perceiving the world as a game of codes and symbols may reduce cognitive rigidity and encourage flexibility, allowing individuals to distance themselves from literal interpretations and explore multiple meanings in their experiences \cite{1}. This aligns with game theory concepts, where understanding rules and strategies leads to better adaptation and resilience to life’s challenges.
Moreover, graphical representation of concepts, such as using graphs to model real or abstract spaces, allows for a visual understanding of relationships and connections between various elements. For example, a graph forming a rectangle with four connected nodes and one interior node can symbolize a physical space or a conceptual structure. Through scaling and rotation operations, this form can be identified at micro or macro levels of reality. This fractal perspective suggests that patterns appear again across different scales, offering a unifying view of the universe’s structure.
Objectives of this theory are:
\begin{itemize}
    \item Developing a theoretical framework that integrates numbering, linguistic coding, and graphical representations to explore connections between different domains of knowledge.
    \item Applying the principles of symbolic theory in therapeutic contexts to provide patients with new perspectives and tools for understanding and managing psychological experiences.
    \item Investigating fractal relationships and recurring patterns in linguistic and mathematical structures to discover innovative ways of perceiving and interpreting reality by identifying the efficient cause.
\end{itemize}
These aspects will be developed in the next sections.
\\
\\
\\
\\
\noindent \LARGE \textbf{II Theoretical Notions}
\\
\\
\large In this section, the main principles will be theoretically analyzed in a way that familiarizes the audience with concepts.
\\
\\
\\
\noindent \textbf{Symbol and Coding}
\\
\\
\noindent Symbols and their coding have long been fundamental tools for understanding reality, with applications in psychology, mathematics, and philosophy. Carl Jung explored symbols as part of the collective unconscious, considering them universal archetypes that shape our perception of the world and influence how we relate to personal experiences, including traumas \cite{2}. In the context of numerology and linguistics, Pythagoras emphasized that numbers are essential for deciphering universal order, while modern research in linguistic semantics confirms that deconstructing words into syllables and assigning numerical values can reveal hidden semantic patterns \cite{lake}. These ideas provide a foundation for symbolic analysis and linguistic coding, offering premises for a fractal understanding of reality.
\\
\\
\\
\noindent \textbf{Fractal Analysis}
\\
\\
Fractality, introduced by Benoît Mandelbrot, demonstrated that natural structures follow recurring patterns regardless of scale. These include large-scale phenomena such as coastlines and microscopic structures like bronchial trees in the lungs \cite{3}. The concept of fractality has been extended to studies on temporal proportionality, such as chronobiology, where time is perceived as a succession of recurring cycles \cite{smale}. These principles can also be applied to linguistic contexts, where fractality may represent the recursive structure of language or the temporal relationships between events, such as dates and times. For instance, the ratio between a day and a month, or between hours, minutes, and seconds within a time interval, can reflect a fractal symmetry that indicates order amidst apparent chaos.
\\
\\
\\
\noindent \textbf{Implications in Psychological Therapy}
\\
\\
\noindent The relationship between symbolic coding and psychological therapy is explored through techniques such as narrative therapy, where externalizing thoughts as symbols or stories allows patients to reframe traumas \cite{white}. Similarly, symbolic numbering of fingers—where each finger can represent an event, emotion, or date—adds a bodily dimension to the reframing process. This approach is supported by recent findings showing that the simple symbolic organization of information can reduce trauma-related anxiety and encourage cognitive flexibility \cite{turn}.
\\
\\
\\
\noindent \textbf{Presence of Game Theory}
\\
\\
\noindent Game theory, developed by von Neumann and Morgenstern, adds a strategic dimension to understanding symbols. By viewing the world as a ``game of codes'', individuals can explore different ways of interacting with reality, treating it as a set of rules that can be learned and utilized \cite{4}. This approach is particularly relevant for OCD, where patients can learn to see obsessions as symbolic codes that can be analyzed and reconfigured. Furthermore, symbolic games can be used to explain connections between fractal proportions and linguistic coding, clarifying the interpretation of reality’s complexity.
\\
\\
\\
\noindent \textbf{Temporal Proportions}
\\
\\
\noindent A fascinating direction is the link between dates and times, where their proportions provide a pathway for deciphering temporal patterns. Research on time’s fractality suggests that relationships between dates and times are not random but reflect subtle patterns of universal order \cite{data}. For example, associating a day like October 15th with a specific time, such as 15:10, can be analyzed through numerical and symbolic ratios. This method has profound implications for understanding cyclical phenomena and therapeutic applications, where temporal anchors can be used to reframe traumatic events.
\\
\\
\\
\noindent \textbf{Gaps}
\\
\\
\noindent While symbolism, fractality, and reality coding have been studied in various ways, there is an evident lack of integration between these fields. Symbols are rarely explicitly connected to fractality, and the coding of language or time has not been sufficiently utilized in therapy or education. The Theory of Proportions and Symbolic Allusions (TPASAI) addresses these gaps through an interdisciplinary approach that coherently connects symbols, fractality, and proportionality. TPASAI not only brings new understanding to the relationships between these concepts but also offers practical applications in cognitive therapy, education, and technology.
\\
\\
\\
\\
\noindent \LARGE \textbf{III Methodology}
\\
\\
\noindent \large TPASAI represents a scientific approach that integrates mathematical proportion analysis, symbolic coding, and fractal principles to explore connections between dates, symbols, and events. This methodology involves identifying and coding symbols, analyzing numerical and fractal relationships between them, and applying these results across diverse domains such as psychology, education, and technology (Fig. \ref{fig:enter}).

The first step in TPASAI involves transforming calendar, temporal, and linguistic data into numerical or symbolic codes for analysis. For instance, the date 11/12 can be proportionally represented as the ratio $\frac{11}{12}$, and the time 15:10 as $\frac{15}{10}$. These values provide a mathematical basis for identifying recurring and proportional patterns. Similarly, words are broken into syllables and roots, with each component numerically coded based on its length or position in the alphabet. For example, the word ``suspect'' is divided into ``sus'' and ``pect'', whose lengths (3 and 4) generate the code (3, 4), later used for symbolic analysis.

The same method applies to symbolic numbering of the body, where fingers are numbered from 1 to 10, each assigned a numeric value. This technique allows the association of dates or events with tangible elements, such as fingers, facilitating proportional analysis and symbolic interpretation.

After coding, the proportional and fractal analysis stage examines the generated numerical ratios to identify patterns and recurring relationships. Relationships between symbols, such as the ratio $\frac{11}{12}$, are extended to compare across different scales and contexts. For instance, the ratio $\frac{11}{12}$ can be correlated with other temporal or numerical proportions, such as $\frac{11}{12}$, to identify symmetries or recurring patterns. In this process, graphical visualizations such as two-dimensional or three-dimensional graphs are used to represent the connections between symbols and proportions for various real-world locations associated with a miniature form of it when rescaling is applied.

The results from these analyses are validated through practical applications. In psychology, TPASAI is used to reframe significant events or traumas by transforming them into positive symbolic structures through numerical coding and proportional analysis. For example, a traumatic event associated with a specific date can be reinterpreted using fractal analysis of temporal and symbolic proportions. In education, the methodology is applied to teach mathematical or linguistic concepts, where students explore connections between calendar dates and relevant events, developing a deep understanding of numerical structures and proportions. In technology, TPASAI is used to develop ciphering keys based on symbolic codes or in natural language processing algorithms.

To implement TPASAI, mathematical software such as Python is utilized to calculate proportions, perform fractal analysis, and generate graphical visualizations. Symbolic networks are constructed to analyze the relationships between numerical and linguistic elements, offering intuitive representations of identified connections. Experimental studies validate the methodology’s effectiveness, involving participants in the proportional and symbolic analysis of personal data. Although interpreting symbols may vary depending on cultural or subjective contexts, the flexibility of the methodology allows adaptation to multiple domains and populations.

Through the integration of symbolic coding, proportional analysis, and fractal principles, TPASAI provides an interdisciplinary framework for discovering recurring patterns and applying them in psychology, education, and technology. The methodology stands out for its ability to combine the rigorous structure of mathematics with practical interpretations, contributing to a deeper understanding of data and events through a proportion- and symbol-based approach.
\\
\\
\\
\\
\noindent \LARGE \textbf{IV Conclusion}
\\
\\
\large The TPASAI methodology offers an innovative framework for understanding and applying symbolic, proportional, and fractal relationships in various contexts. By integrating symbolic coding, proportional and fractal analysis, and validation through practical applications, TPASAI demonstrates a structured and interdisciplinary approach. It finds applicability in psychology for reframing traumas, in education for deepening conceptual understanding, and in technology for ciphering and natural language processing.

The methodology’s graphical representation clearly illustrates the stages and their interconnections, emphasizing the interdependence between identifying symbols, mathematical analysis, and validation through practical applications. Adjustments made for clarity and readability in organizing the information contribute to a better understanding of the methodology.

By combining rigorous mathematical structure with applied interpretations, this methodology has the potential to open new directions for interdisciplinary research and development, promoting a deeper understanding of events and relationships in real-world actions and behaviors.

$\,$

$\,$

\end{document}